\documentstyle[psfig]{l-aa}

\begin{document}

\thesaurus{
       (12.04.3;  
        12.12.1;  
        11.03.1;  
        11.12.2)  
          }

\title{Cluster luminosity function and n$^{th}$ ranked magnitude as
a distance indicator 
       \thanks{Based on observations collected at the European Southern
               Observatory (La Silla, Chile)}
       \thanks{http://www.astrsp-mrs.fr/www/enacs.html}
      }

\author{ S.~Rauzy \inst{1}, C.~Adami \inst{2}, A. Mazure \inst{2}}
\institute{ Centre de Physique Theorique, Marseille, France 
   \and IGRAP, Laboratoire d'Astronomie Spatiale, Marseille, France }

\offprints{C.~Adami}
\date{Received date; accepted date}

\maketitle
\markboth{Cluster luminosity function and n$^{th}$ ranked magnitude as
distance indicator}{}

\begin{abstract}

We define here a standard candle to determine the distance of 
clusters of galaxies and to investigate their peculiar velocities by using the 
n$^{th}$ rank galaxy (magnitude m$_n$). We address the question of 
the universality of the luminosity function for a sample of 28 rich clusters of 
galaxies (cz$\simeq$20000 km.s$^{-1}$) in order to model the influence on 
m$_n$ of cluster richness. This luminosity function is found to be universal 
and the 
fit of a Schechter profile gives $\alpha$=-1.50$\pm$0.11 and M$_{bj}*$=-19.91
$\pm$0.21 in the range [-21,-17]. The uncorrected distance indicator m$_n$ 
is more efficient for the first ranks n. With n=5, we have a dispersion of 
0.61 magnitude for the (m$_n$,5log(cz)) relation. When we correct for the 
richness effect and subtract the background galaxies we reduce 
the uncertainty to 0.21 magnitude with n=15. Simulations show that a large 
part of this dispersion originates from the intrinsic scatter of the standard 
candle itself. These provide upper bounds on the amplitude $\sigma_v$
of cluster radial peculiar motions. At a confidence level of 90 $\%$, the
dispersion is 0.13 magnitude and $\sigma_v$ is limited to 1200 km.s$^{-1}$ for 
our sample of clusters.

\end{abstract}


\section{Introduction}

Distances of clusters of galaxies are obtained by measuring the redshift of the 
galaxies inside the clusters. However, some local mass concentrations could 
induce peculiar motions superimposed to the Hubble flow (e.g. Bahcall \& Oh 
1996). Measurements of these peculiar motions which have important consequences
on cosmological models, require the use of independent distance estimates.

\begin{figure*}
\vbox
{\psfig{file=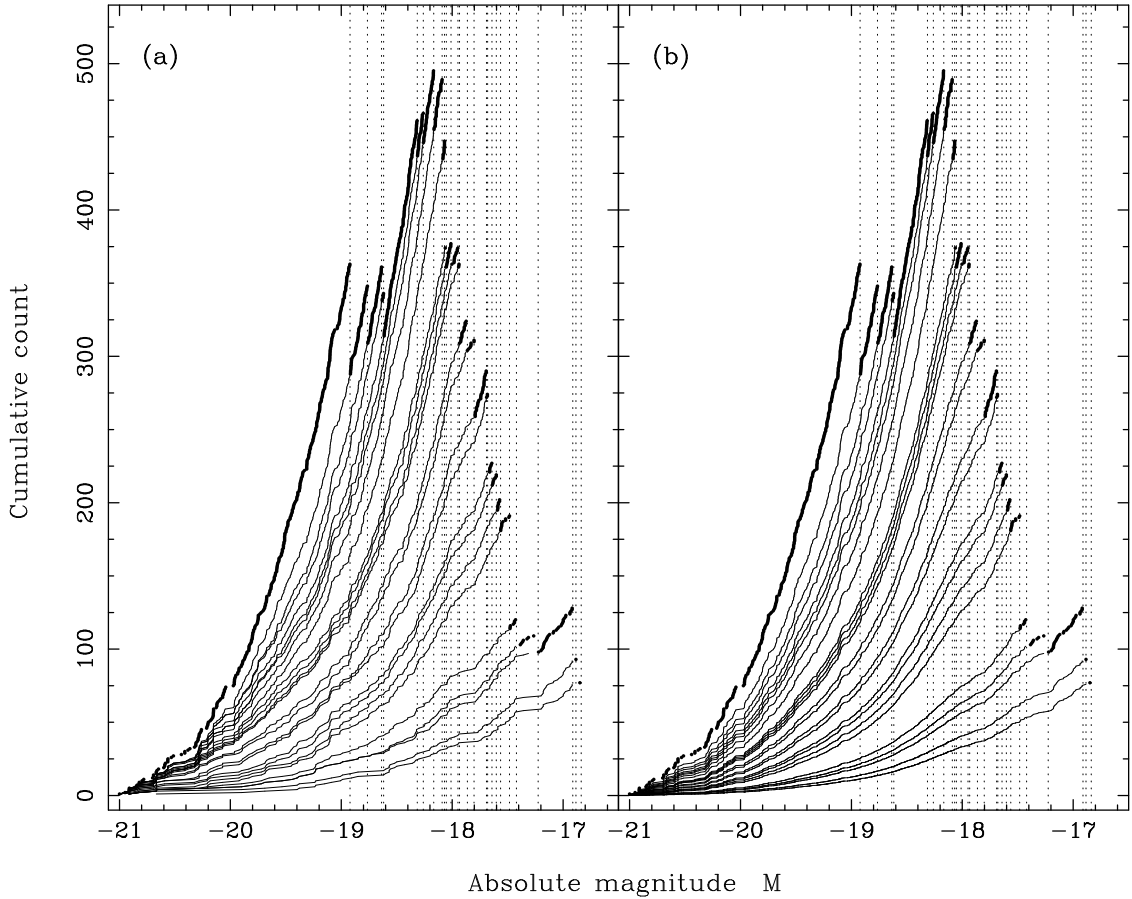,width=20.0cm,angle=270}}
\caption[]{Illustration of the reconstruction method (see text in Sec. 3.1.
for explanations) of the individual and composite luminosity functions.}
\label{un}
\end{figure*}

One of the methods uses the Fundamental Plane (FP hereafter) of clusters
of galaxies (e.g. Schaeffer et al. 1993, for 16 clusters with a median 
redshift of 0.04). Adami et al. (1998: A98a hereafter, for 29 clusters with a 
median redshift of 0.07) show in this way a limit for the peculiar cluster 
motions of less than 1000 km.s$^{-1}$. 
However, the use of the FP as a distance indicator is not easy: the 
determination of the total luminosity requires highly accurate photometry and 
the determination of the core radius requires the positions of the members 
galaxies. 
Finally, to have a reliable velocity dispersion, more than 10 redshifts 
(interlopers removed) are needed. It is also common to use the tenth rank 
galaxy as a standard candle (e.g. Abell, Corwin \& Olowin 1989: ACO 
hereafter) to find the distance. Bahcall \& Oh (1996) use  a sample of cluster 
velocities based on Tully Fisher distances of Sc galaxies for the same goal. 
Colless (1995), Hudson \& Ebeling (1997) or Lauer \& Postman (1994) use also 
the slope of the brightness profile of the cD galaxies to deduce the distance. 
Using these different methods, these authors constrain the peculiar velocities
of different samples of nearby clusters (z$\leq $0.05).

To re-address these questions, we develop in this work a new distance indicator
using the galaxy of the n$^{th}$ rank for a given cluster. We use the ENACS 
(see Katgert et al. 1996, Mazure et al. 1996, Biviano et al. 1997, A98a, 
Katgert 
et al. 1998 (K98), Adami et al. 1998 (A98b) and de Theije et al. 1998) and 
COSMOS (e.g. Heydon Dumbleton et al. 1989) data. To estimate correctly 
this indicator, we look at the possible universality of the Luminosity Function
for clusters of galaxies (LF hereafter), after taking into account the
correction for parameters such as the number of galaxies 
in the cluster. The universality of the LF is for example treated in Lumsden 
et al. (1997: L97 hereafter), in Valotto et al. (1997: V97 hereafter) or in
Trentham (1998). L97 and V97 
derive synthetic LF's for samples of clusters of galaxies. V97 found a 
significant difference between the rich and the poor clusters, while L97 found 
a significant difference between the high and low velocity dispersion clusters.
These two studies use the COSMOS/EDSGC data. For V97, the redshifts of the 
clusters come from a literature compilation.

In order to study these points and to investigate the existence of a standard
candle, we will proceed as follow: In the first section we describe the 
selected sample. We determine in the second section the luminosity function 
and we look for its universality, at least in the magnitude range [-21;-17]. 
In the third section, we redefine m$_n$ as a distance indicator and correct it 
for the influence of background level, the richness of the clusters and the
statistical effects and we use it to examine the peculiar motions of the 
ENACS/COSMOS clusters of galaxies.

We use in this article H$_0$=100 km.s$^{-1}$.Mpc$^{-1}$ and q$_0$=0.

\section{The Sample}

In this work, we use the ENACS and COSMOS surveys. They are well described in 
K98.
ENACS gives the redshifts and the R$_{25}$ calibrated magnitude and COSMOS the 
positions and the b$_j$ magnitudes for all galaxies in the clusters. K98 show 
that clusters for which the photometry in the two catalogues is based on 
the same survey plate, the two magnitude scales agree very well. There do not
appear to be serious problems with either magnitude scale. 
In addition, some redshifts come from the literature. The
absolute magnitudes have been computed by using the mean cluster redshift and
the same K(z)-correction as in L97: K(z)=4.14z-0.44z$^2$. We have also
corrected for galactic extinction using the map of Burstein \& Heiles (1982)
in the same way as in A98a. The redshifts are calculated with respect to the 
rest frame of the Cosmic Microwave radiation
(CMR hereafter) defined by Lubin \& Villela (1986).To 
reduce the substructure effects and to have good measurements of the different
cluster parameters such as core radius, velocity dispersion, mean redshift,
background level and number of galaxies on the line of sight, we limit the 
global sample to the 29 most regular clusters in 
A98a. These clusters have an Abell richness greater than 1. 
They do not have major 2D visible substructure. We take only the unatypical 
King core radii (we have removed A3128 which exhibits a large core radius). 
For these final 28 clusters we consider galaxies within 5 King core radii 
(about 500 kpc). 

According to K98, we know that COSMOS has a completeness level of 91\% for 
b$_j$ $\leq $19.5. However, this estimate refers to areas with high surface
density in the COSMOS catalogue. For low surface density of galaxies (outside
the clusters), the completeness level in certainly higher. To increase the
completeness, we add to the COSMOS objects, the ENACS ones not found in the 
given area and inside the clusters. Finally, to be sure that we have a 
complete sample, we limit it to b$_j$ $\leq $19. We note also that we remove 
the objects with a back- or fore-ground redshift (see Katgert et al. 1996). We 
have finally more than 
3500 galaxies in the global sample. We split the sample into 3 sub-samples to 
test for spatial variations. The first sub-sample contains the galaxies 
between 0 and 2 core radii, the second the galaxies between 2 and 3.5 core 
radii and the third the galaxies between 3.5 and 5 core radii. We have almost 
1200 galaxies in each of the three sub-samples.

\section{The cluster luminosity function}

\subsection{The method}
In order to test the universality of the LF for our 28 clusters, we construct 
from the present data this function. We use a method similar to those used for 
example in Beers $\&$ Tonry 
(1986) or Merrifield \& Kent (1989) for the density profiles reconstruction. 
We take into account the different limiting magnitudes of the different 
clusters. We consider the composite cumulative LF (CCLF hereafter, 
noted F(M) in the figures). L97 use a similar (Colless 1989) reconstruction 
method, while V97 simply add the individual clusters with a common limiting 
apparent magnitude of 19.4.

First of all, we remove statistically the background objects in each
cluster. The mean number of removed objects is the mean number of
background objects minus the number of already removed objects 
on the basis of the redshift. The mean number of background objects comes 
from a density profile fit, including as a free parameter the background 
density (A98a and b ). A98a and b have shown that this background estimation
is very robust and in good agreement with the count law of field galaxies. 
It represents about 44 $\%$ of the total number of 
galaxies along the line of sight. Starting from a limiting magnitude 
of $b_j$=20, 
we have rescaled this number for $b_j$=19 (for better completeness) by using 
the count law of L97 in order to calculate the proportion of galaxies in each 
magnitude 
bin. We also use this law to select the magnitude of the background removed 
objects. To obtain a statistical error, we have made 100 calculations of the
cumulative LF for each cluster (CLF hereafter), taking into account the 
internal background fluctuation (i.e. the error in the determination of the 
mean number of background objects for each cluster). 

As described above, we have applied to construct the CCLF an adapted 
version of the method devised for example by Beers and Tonry (1986) for the 
cluster density profiles. 

We denote by $M_{lim}^k=m_{lim}^c-\mu(z)$ the corrected absolute 
magnitude limit of the $k^{th}$ farthest clusters. Up
to each $M_{lim}^k$ we compute the cumulative
count $G_k(M)$ of all the galaxies belonging to the set
of clusters $i$ complete in $M$ i.e.  
verifying $M_{lim}^i \le M_{lim}^k$ (hereafter these 
galaxies samples are called $S_k$). While these
cumulative counts $G_k(M)$ are not affected by incompleteness
problems, they suffer from sampling errors
as $k$ increases (i.e. because the number of selected clusters
decreases with distance, the number of galaxies in sample 
$S_k$ for a given absolute magnitude range is a decreasing
function of $k$). Fig. 1(a) illustrates this behaviour:
the $G_k(M)$ of samples $S_k$ are plotted. The cut-off
$M_{lim}^k$ are indicated as dotted lines. The number of clusters contributing
to a $G_k(M)$ decreases with $M_{lim}^k$. In order to minimize sampling 
errors, we adopt the
following rescaling procedure for reconstructing the CCLF.
For an increasing cut-off $M_{lim}^k$,
rescaled cumulative counts $F_{k+1}(M)$ are
defined recursively as follows
$$
F_{k+1}(M) = \frac{G_{k+1}(M_{lim}^k)}{F_k(M_{lim}^k)} F_k(M)
~~{\rm if}~~ M \le M_{lim}^k 
$$
$$
F_{k+1}(M) = G_{k+1}(M)
~~{\rm if}~~ M_{lim}^k < M \le M_{lim}^{k+1}
$$
with $F_{1}(M)=G_{1}(M)$. It consists in replacing up to $M_{lim}^k$ 
the cumulative counts $G_{k+1}(M)$ by the reconstructed CCLF
$F_k(M)$ renormalized such that continuity of the final CCLF is 
ensured. The sampling errors
are thus minimized ($F_k(M)$ are plotted Fig. 1(b) for comparison) since
$F_k(M)$ is estimated using information provided by all the sampled
clusters while cumulative count estimate $G_{k+1}(M)$ use only the set of
$i^{th}$ clusters verifying $M_{lim}^i \le M_{lim}^{k+1}$. Such a procedure
warrants an optimal reconstruction of the
cumulative luminosity function of galaxies belonging to a sample of clusters
spread in redshifts.

\begin{figure}
\vbox
{\psfig{file=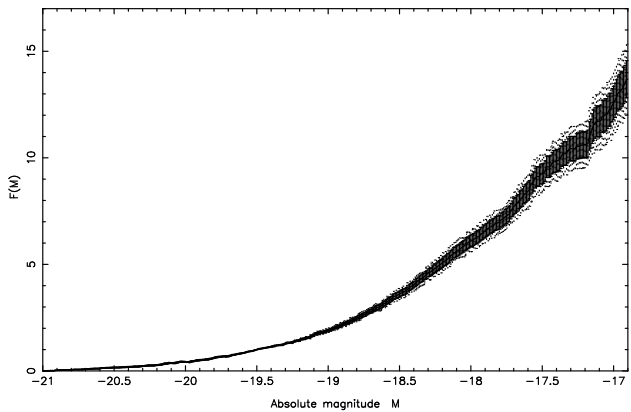,width=9.0cm,angle=270}}
\caption[]{Final CCLF with errors. The filled surface
represents the envelope (the error) of the CCLF due to the field subtraction 
and the dotted envelope takes into account the statistical 
fluctuations. The CCLF is normalized to 1 for
M=-19.5.}           
\label{deux}
\end{figure}

The final CCLF is shown in Fig.2, arbitrarily normalized at 1 for 
M=-19.5. It spans a range of 4 absolute magnitudes M between -21 and -17. The
lower limit allows us to exclude the very bright galaxies (cD galaxies) which 
are probably not belonging to the mean luminosity function. The upper limit 
corresponds to the absolute magnitude of the faintest galaxy with m$_{lim}$=19
in the nearest cluster. L97 have used the range [-21;-18] and V97
the ranges [-21.5;-17] and [-21.5;-16].

This method assumes obviously that the LF's are similar in the different
clusters. We will check afterwards that this condition is
well satisfied (see Sec.~3.2.3).

\subsection{Analysis}

\begin{figure}
\vbox
{\psfig{file=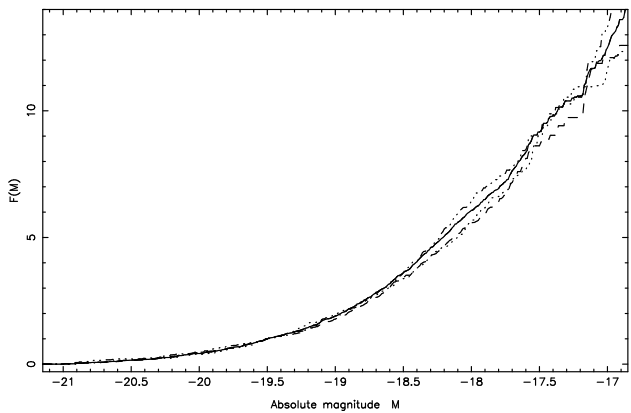,width=9.0cm,angle=270}}
\caption[]{CCLF's issued from the three radial bins: [0,2] r$_c$ (filled), 
[2,3.5] r$_c$ (dashed) and [3.5,5] r$_c$ (dashed dotted).}
\label{trois}
\end{figure}

\subsubsection{Spatial variations}
We looked for spatial variations of the LF. In order to do that, we 
tested the CCLF in the 3 defined areas. We reconstructed it exactly in the same 
way as above but for zones enclosed in [0,2] r$_c$ (CCLF1), [2,3.5] r$_c$
(CCLF2) and [3.5,5]r$_c$ (CCLF3). We superpose the three CCLF's (Fig. 3) after
normalization at 1 for M=-19.5. They look very similar. A Kolmogorov-Smirnov 
show that CCLF1 and CCLF2 are identical at a confidence level of 95 \%.
CCLF2 and CCLF3 are identical at a confidence level of only 65\%. The 
conclusion is that the 3 CCLF's are
probably identical (with a high confidence level) and that the LF's do not 
vary significantly with the radius. This result is in agreement with V97, which
have used 55 clusters (16 rich clusters). Only 3 of those clusters are common
with our sample. They did not find significant variations of the LF within 1 
Mpc from the center (roughly 10 core radii).

Moreover, this universality supports our background subtraction approach. We 
assumed 
implicitely that the background is homogeneous inside 5 core radii. The
number of removed galaxies in each of the three bins is then directly
proportional to the area of these bins. We have thus removed 4 times more 
galaxies in the exterior bin than in the central one.  However,
the three reconstructed CCLF's are very similar, supporting the way we remove
the background.

\subsubsection{Universality of the LF}

A way to test this point is to compare the individual clusters CLF to the
CCLF (Fig.~4). We plot all the individual CLF's in apparent magnitudes and no 
subtracted background (G(m) in Fig.~4). We 
simulate the theoretical no background subtracted CLF's 
by adding background objects (L97) to the reconstructed mean
CCLF (normalized to the number of galaxies in each cluster). We see
a good agreement between the observations and the simulations. We can quantify
this agreement by using a Kolmogorov-Smirnov test between the observed data 
and the simulated data. We test the hypothesis that the observations and 
simulations are drawn from the same parent population. 
The mean risk is 62\% for 80\% of the clusters which is a conclusive 
statement: we have a small dispersion of the individual CLF's around the mean 
function, and so the LF is probably universal.

If the individual CLF's are drawn from a universal function, the 
differences (and so the risks) must be randomly distributed. In order to test 
this 
hypothesis, we generate 500 random distributions of 28 CLF's (normalized like 
each real cluster) around the reconstructed global CCLF.

We proceed with another 
Kolmogorov Smirnov test and we find a level of 75\% to reject the right 
hypothesis if we assume that the individual risks distribution is non random.
As a conclusion, we can say that the CCLF's are globally universal for all our 
28 selected rich clusters. This is in agreement with L97 or Trentham (1998). We 
note that V97 argue against a universal LF, but between the rich and the poor 
clusters.

\subsubsection{Modelisation}
Even if the following sections do not use this modelisation, we fit here
a Schechter function (1976) to the LF: 

\[
S(M)=\Phi _{*}10^{0.4(\alpha +1)(M_{*}-M)}\exp [-10^{0.4(M_{*}-M)}] 
\]

where $\Phi _{*}$ is given by the number of galaxies in each cluster and 
$M_{*}$ is the characteristic magnitude.
We use the minimization algorithm MINUIT (e.g. A98a).
First, we calculate a $\chi ^2$ fit using the weights of the 
reconstruction. We have $\alpha $=-1.50 $\pm$ 0.11 and M$_{bj}*$=-19.91 
$\pm$ 0.21. The LF's calculated with the 3 radial zones give similar results 
at the 1 sigma level. 
If we minimize the maximal distance (divided by weight) between the model 
and the observations instead of the $\chi ^2$ (Kolmogorov Smirnov fit: KMS fit
hereafter), we have $\alpha $=-1.47 and M$_{bj}*$=-19.89 (without reliable 
errors).

These two results are consistent at the 1 sigma level (according to the error 
bars). 
Morever, we have a difference of less than 2 $\%$ for $\alpha $ and 1 
$\%$ for M$_{bj}*$. The parameter determination is then independent of the
fitting method.

These values could also be compared with the recent analyses of L97 and V97.
They have both used the $\chi^2$ minimization.

V97 have found $\alpha $=-1.5$\pm $0.1 and M$_{bj}*$=-20.0$\pm $0.1 for their 
rich clusters and for the magnitude range [-21.5;-17]. Those two parameters 
are in agreement at the 1 sigma level with our KMS or $\chi^2$ 
determinations.

L97 find $\alpha $=-1.22$\pm $0.04 and M$_{bj}*$=-20.16$\pm $0.02 in the 
magnitude range [-21,-18]. The result for M$_{*}$ is also in good agreement 
with our value at the 1 sigma level. However, the $\alpha$ value is only 
consistent at the 3 sigma level. If we fit into the same magnitude range we 
find M$_{bj}*$=-20.18$\pm $0.20 in perfect agreement with L97, but 
$\alpha $=-1.63$\pm $0.12 consistent at only 3 sigma. We note that L97 use 
q$_0$=1, but it has no influence on the M$_*$ determination.

For individual clusters, we compare with Bernstein et al. (1995) and Lobo
et al. (1997). They found respectively $\alpha $=-1.42$\pm $0.05 and $\alpha 
$=-1.59$\pm $0.02 for the Coma cluster. The two values are consistent with our 
KMS fit at the 1 
sigma level. We note here that these two values are deduced from photometric 
surveys of the core of the Coma cluster after statistical background 
subtraction. 
For the Lobo et al. study, we have considered the result for a Schechter 
profile fit only.

Nearly all the results cited in the literature are consistent with 
our parameters. The only discrepancy occurs for the $\alpha$ value of 
L97. This could be due to two major sources: 

-First, Bernstein et al. (1995), L97, V97 or Lobo et al. (1997) remove the
background galaxies in a statistical way. The most local corrections are made 
in L97 and consist in the removal of a uniform background density calculated 
in an external annulus. But, the radius of this annulus is always greater than 
4 Mpc. We show in A98a that the background density may change by a 
significant factor at these scales. As an example, the two clusters A3825 and 
A3827 are separated only by about 3 Mpc and the background density for A3825 is 
40 $\%$ higher than for A3827. Removing the background by using a distant 
external annulus could then induce a bias.  

-Second, L97 use all the COSMOS galaxies brighter than b$_j$=20. We know (see 
K98) that the COSMOS catalogue in the area of ENACS clusters is only complete 
at the 90 $\%$ level for b$_j$$\leq$19.5 and we limit here our sample to
b$_j$$\leq$19 to be sure to be complete. The LF's of L97 could then miss some 
galaxies in the faint parts, which could lead to a lower $\alpha$ value.

\begin{figure*}
\vbox
{\psfig{file=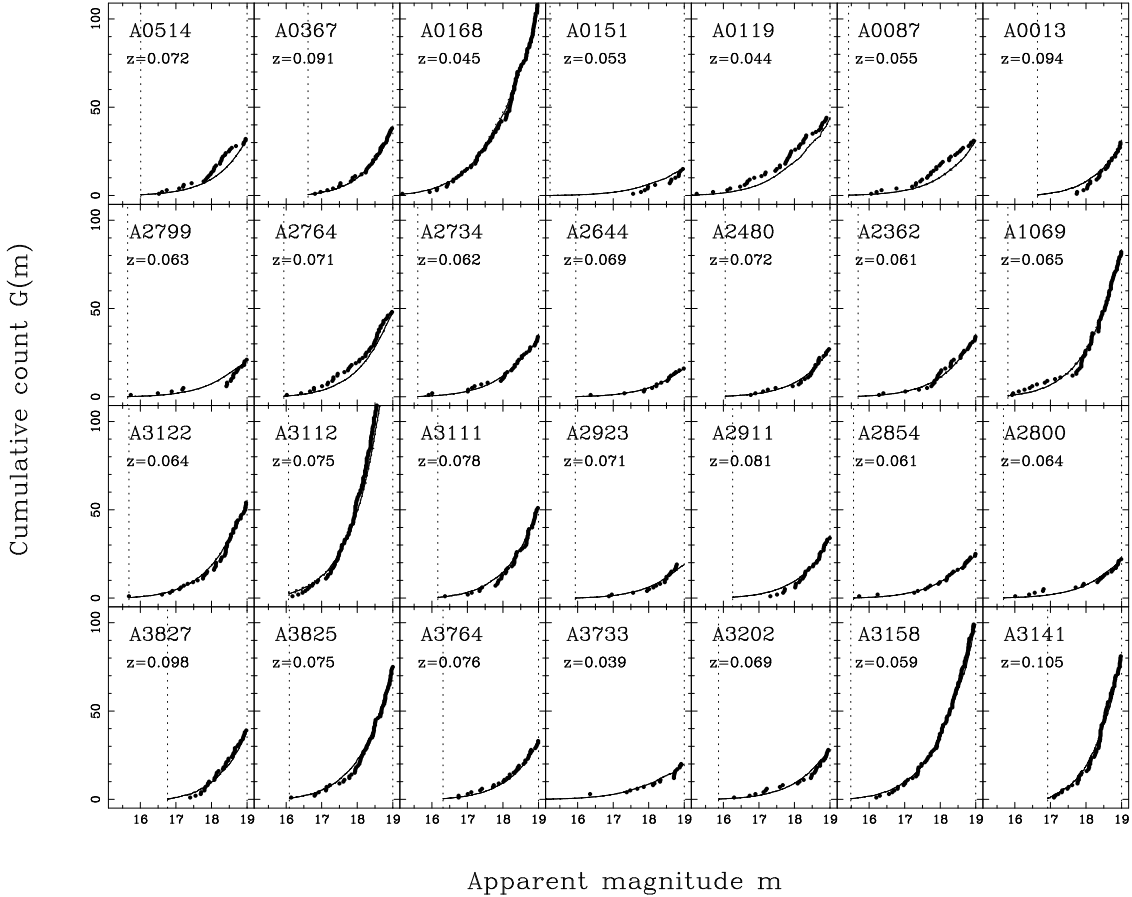,width=20.0cm,angle=270}}
\caption[]{The 28 observed CLF's + background (points). The model CLF's deduced
of the CCLF after normalization + background are superposed (solid line).}
\label{quatre}
\end{figure*}

\section{The distance indicator}

After we have shown that LF's are universal within the considered
magnitude range, we want to test the Hubble flow and to try to determine 
the peculiar velocities in our sample of clusters of galaxies by using a 
distance indicator using photometric data.

\subsection{Definition of the optimal rank n for the distance determination}

Following Jones \& Mazure (1993), who have used $m'_{15}=\frac 1{11}\sum
\limits_{r=10}^{20}m_r$ (where m$_r$ is the magnitude of the r$^{th}$ ranked 
cluster galaxy), we search for a similar standard candle. We test here m$_n$
with n$\in $[1,28]. We first look at the
Hubble relation for the m$_{10}$ magnitude to be
coherent with ACO (1989). By using bisector indicators (Isobe et al., 1990),
we find a regression slope between m$_{10}$ and log10(cz) of 4.58$\pm $1.24
consistent with the previous ACO results. This is in good
agreement with the expected value of 5. So, we will fix hereafter the slope of 
all the regressions between m$_n$ and log10(cz) to a value of 5.

We want now to find the optimal rank to define a distance indicator. To deal
with the real minimal observing conditions, we use all the projected
galaxies: we do not remove from the total sample the fitted number of 
background galaxies. We
compute the basic dispersion of the relations between m$_n$ and log10(cz) 
for each n. The best choice is the rank n for which the dispersion is 
minimal. We see on Fig.~5 that the basic dispersion is minimal for the 
first ranks. If we take for example n=5 (see tab.~1), the dispersion 
is 0.61 
magnitude. If we assume as the mean velocity of our cluster sample 
20000 km.s$^{-1}$, the corresponding dispersion in velocity is 5612 km.s$^{-1}$
(0.02 in redshift).
We have m$_5$=5log(cz)-(4.46$\pm$0.12).

Clearly, this precision is too low to allow any analysis of the peculiar
velocities of the individual clusters. 

\subsection{Peculiar velocities}

The indicator used previously is affected by different intrinsic factors
peculiar to each cluster, such as the background level, the richness of
the clusters (number of galaxies inside these clusters) and of course the
peculiar velocities. We want here to correct for the background and for the 
richness. To model these two contributions, we assume first that in our 
absolute magnitude range [-21.,-17.], the count law of the background galaxies 
G$_b$(m) is proportional to the canonical exponential: 
exp($\alpha$(m-m$_{lim}$)) with $\alpha$ the logarithmic slope and m$_{lim}$ 
the apparent limiting magnitude (=19.). Second, we note F$_0$(M) (with M the 
absolute magnitude) the CLF normalized to unity at M$_0$=-19.5. If $\mu$ is 
the distance modulus of a given cluster, we have M=m-$\mu$.
We deduce then an expression for the rank k of a galaxy in a given cluster with
N$_c$ member galaxies and N$_b$ background galaxies (according to the
apparent limiting magnitude m$_{lim}$=19):

k=N$_c$F$_0$(M$_k$)+N$_b$G$_b$(m$_k$)=N$_c$F$_0$(m$_k$-$\mu$)+N$_b$G$_b$(m$_k$)

We derive then the corrected value of the k$^{th}$ magnitude:

m$_k$-$\mu$=F$_0$$^{-1}$((1/N$_c$)(k-N$_b$G$_b$(m$_k$)))

where N$_b$ is deduced from the fits of the different density profiles (see 
A98a) and N$_c$ is the observed individual number of cluster galaxies with 
M$\leq$M$_0$.

\begin{figure}
\vbox
{\psfig{file=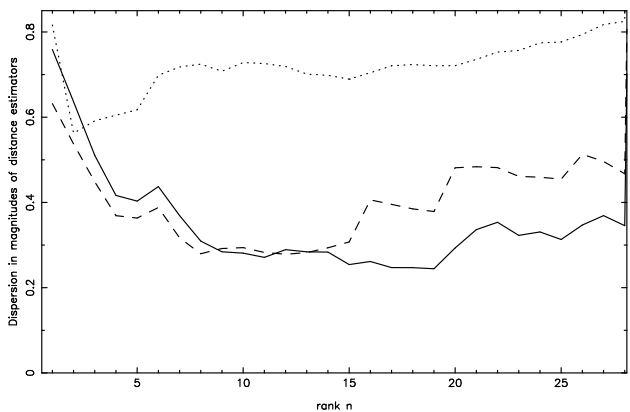,width=9.0cm,angle=270}}
\caption[]{Dispersions in magnitude according to the rank k. The dotted line is
the first indicator without correction, the dashed line is the indicator
corrected for richness and the solid line is the indicator corrected for
richness and for background galaxies.}
\label{cinq}
\end{figure}

\begin{figure}
\vbox
{\psfig{file=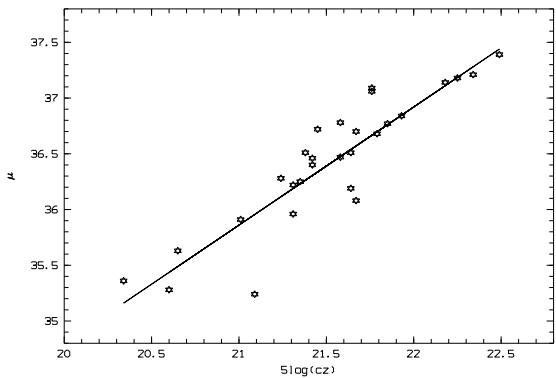,width=8.0cm,angle=270}}
\caption[]{($\mu$-5log(cz)) relation with the distance indicator corrected
for the richness and for the background and calculated with m$_{15}$. 
The cluster with the lower $\mu$ is A0087.}
\label{six}
\end{figure}

We see in Fig.~5 the improvement resulting from the corrections.
We note that the dispersion does not decrease after the 20$^{th}$ rank
because we start to deal with clusters with less than 20 galaxies in the 
studied areas. We note also that a correction for background galaxies using 
mean densities instead of our local estimation is not very efficient. The 
final dispersion of 0.254 leads to a precision in velocity of about 2300 
km.s$^{-1}$. We show in Fig.~6 the relation between the corrected distance 
modulus $\mu$=m$_{15}$-M$_{15}$ of each of the 28 clusters and 5log(cz) (see 
also Tab.~1). m$_{15}$ is the 
measured apparent 15$^{th}$ magnitude and M$_{15}$ is the 
absolute magnitude corrected for richness and background effects.
Removing A0087 from the sample (the atypical cluster in Fig.~6), we obtain a
dispersion of 0.210 magnitude (1900 km.s$^{-1}$). Durret et al. (1998) 
argue that A0087 is not really a cluster, but the result of a superposition 
effect. A part of the dispersion is due to statistical
fluctuations originating from finite sample size effects when we reconstruct 
the CCLF (see $\S$ 3.1). We have quantified this effect by carrying out 1000 
Monte-Carlo simulations.
We estimate then the probability P($\sigma_{stat}\leq\sigma$) to have a 
statistical contribution lower than a given value $\sigma$. According 
to the observed value of the magnitude dispersion $\sigma_{obs}$ and assuming 
that all the remaining dispersion (corrected for richness, background and 
statistical effects) is due to peculiar motions, we can deduce the probability 
to have an error if we assume that the remaining dispersion is lower than 
($\sigma_{obs}^2$-$\sigma^2$)$^{1/2}$. We are thus able to give an upper
limit for the amplitude of the cluster radial peculiar motions at a given 
confidence level. 
We compute these equivalent upper 
peculiar motions by adopting a mean velocity of 20000 km.s$^{-1}$ (see Fig.~7).
If we consider for example a risk of 45 $\%$ for P($\sigma_{stat}\leq\sigma$),
we give a dispersion of 0.11 magnitude equivalent to peculiar motions within 
an amplitude of 1000 km.s$^{-1}$ (0.09 magnitude or 800 km.s$^{-1}$ without 
A0087). With a conservative risk of 10$\%$, we predict a dispersion of 0.17
magnitude equivalent to peculiar motions less than 1500 km.s$^{-1}$ (0.13
magnitude or 1200 km.s$^{-1}$ without A0087).

\begin{figure}
\vbox
{\psfig{file=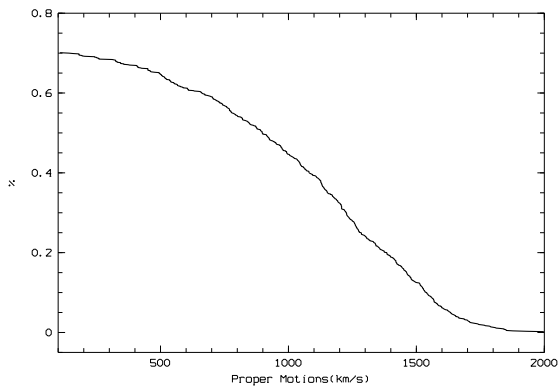,width=8.0cm,angle=270}}
\caption[]{Probability of having peculiar velocities greater than a given 
value (we 
have adopted a mean velocity of 20000 km.s$^{-1}$ for our clusters). This 
curve is shifted of about -300 km.s$^{-1}$ if we remove A0087 from the sample.}
\label{sept}
\end{figure}

\begin{table} 
\caption[]{Parameters for the 28 used clusters. Col.(1) cluster name, 
col.(2) m$_5$ without any corrections, col.(3) distance indicator $\mu$
corrected for richness and background galaxies and calculated with
m$_{15}$, col.(4) 5.$\times$log(cz)}
\begin{flushleft} 
\begin{tabular}{crrr} 
\hline
\noalign{\smallskip} 
cluster name & m$_5$ & $\mu$ &  5log(cz) \\ 
\hline
\noalign{\smallskip}
0013 & 17.64 & 37.18 & 22.25 \\ 
0087 & 16.88 & 35.24 & 21.09 \\ 
0119 & 15.88 & 35.28 & 20.60 \\
0151 & 17.76 & 35.91 & 21.01 \\ 
0168 & 16.23 & 35.63 & 20.65 \\
0367 & 17.05 & 37.14 & 22.18 \\ 
0514 & 16.86 & 36.08 & 21.67 \\ 
1069 & 15.92 & 36.72 & 21.45 \\
2362 & 17.25 & 35.96 & 21.31 \\ 
2480 & 17.09 & 36.70 & 21.67 \\ 
2644 & 17.72 & 36.47 & 21.58 \\
2734 & 16.77 & 36.25 & 21.35 \\ 
2764 & 16.49 & 36.19 & 21.64 \\ 
2799 & 16.95 & 36.51 & 21.38 \\
2800 & 16.53 & 36.40 & 21.42 \\ 
2854 & 17.33 & 36.22 & 21.31 \\ 
2911 & 17.39 & 36.84 & 21.93 \\
2923 & 17.66 & 36.51 & 21.64 \\ 
3111 & 16.80 & 36.77 & 21.85 \\ 
3112 & 16.23 & 37.06 & 21.76 \\
3122 & 16.84 & 36.46 & 21.42 \\ 
3141 & 16.40 & 37.39 & 22.49 \\
3158 & 16.41 & 36.28 & 21.24 \\ 
3202 & 16.91 & 36.78 & 21.58 \\ 
3733 & 16.94 & 35.36 & 20.34 \\
3764 & 16.84 & 36.68 & 21.79 \\ 
3825 & 16.79 & 37.09 & 21.76 \\ 
3827 & 17.29 & 37.21 & 22.34 \\
\hline	   
\end{tabular}
\normalsize
\end{flushleft}
\end{table}

\subsubsection{Analysis}

We have reduced the dispersion of the m$_n$-cz relation by subtracting the 
background galaxies and by modeling the effects of 
the cluster population. We determine in this way an upper limit for the 
dispersion due to peculiar velocities. We deduce that this maximal dispersion 
is somewhat larger than the value of Bahcall \& Oh (1996). They found 
a risk inferior to 5\% to have clusters of galaxies with a random peculiar 
velocity greater than 600 km.s$^{-1}$. However, their studied clusters (and 
groups) have a recession velocity less than 10000 km.s$^{-1}$ and the
equivalent magnitude dispersion is similar to our result.

Several other studies have analyzed the proper motions of the nearby clusters 
in the CMB frame: Lauer \& Postman (1994), Colless (1995) or more recently
Hudson \& Ebeling (1997). They use the slope of the brightness profiles of cD 
galaxies as distance indicator. The Lauer \& Postman (respectively Colless
and Hudson \& Ebeling) distance indicator precision is 0.24 magnitude 
(respectively 0.24 and 0.41 magnitude). This is similar to our precision.

We can also directly compare our dispersion of 0.21 magnitude (without 
statistical correction, see above) with the value of Perlmutter et al. (1997). 
They have used a sub-sample of 28 distant type Ia supernovae to constrain the 
cosmological parameters, and they obtain a dispersion of 0.19 magnitude, very 
consistent with our value. Finally, we have similar values (slightly greater) 
for the peculiar velocities than in A98a with almost the same sample. 

\section{Conclusion}

We have readressed the question of the determination of a distance indicator by
using as standard candle the n$^{th}$ ranked galaxy. 

In order to correct the magnitudes for different factors, we have addressed 
the 
question of the universality of the Luminosity Function for rich clusters 
of galaxies. We have constructed a CCLF by using 28 rich clusters in 
the magnitude range [-21;-17]. The fit of a Schechter model gives 
$\alpha$=-1.50$\pm$0.11 and M$_{bj}*$=-19.91$\pm$0.21 in good 
agreement with other literature results. This function is found to be 
universal for these clusters, consistent with the L97 study.

We have found that the uncorrected distance indicator m$_n$ is more 
efficient for 
the first ranks n. With n=5, we observe a dispersion of about 0.6 magnitude,
too large however to derive correct peculiar velocities.

We then use the CCLF to model the effect of the cluster richness on m$_n$ 
in order to have a better precision and to better constrain the cluster 
peculiar velocities. We correct first for the richness effect and second for 
the 
background galaxies subtraction. This allows to reduce the dispersion to 0.254
magnitude (0.210 without A0087). If we assume that this error is only due to
peculiar velocities, they are 2300 km.s$^{-1}$ (1900 km.s$^{-1}$
without A0087) for a cluster at 20000 km.s$^{-1}$. 
However, a large part of this dispersion is due to statistical effects. By 
using extensive
simulations, we give the probability distribution to have a peculiar motion 
lower than a
given value. For example, with a risk of 10 $\%$, we predict a value of 1500 
km.s$^{-1}$ (1200 km.s$^{-1}$ without A0087).

These results agree well with local estimates. We have also consistent 
results with A98a who used the Fundamental Plane for the same clusters 
as used here.

\begin{acknowledgements}

{We wish to thank Dr. J. Lequeux and R. Malina for a careful reading of the 
manuscript.}

\end{acknowledgements}

\vfill

\begin{thebibliography}{99}

\bibitem{} Abell G.O., Corwin H.G., Olowin R.P., 1989, ApJS 70, 1 (ACO)

\bibitem{} Adami C., Mazure M., Biviano A., Katgert P., Rhee G., 1997, A\&A
331, 439 (A98a)

\bibitem{} Adami C., Mazure M., Katgert P., Biviano A., 1998, A\&A
submitted (A98b)

\bibitem{} Bahcall N.A., Oh S.P., 1996, ApJ 462, L49

\bibitem{} Beers T.C., Tonry J.L., 1986, ApJ 300, 557

\bibitem{} Bernstein G.M., Nichol R., Tyson J.A., Ulmer M.P., Wittman D.,
1995, AJ 110, 1507

\bibitem{} Biviano A., Katgert P., Mazure A., et al., 1996, A\&A 321, 84

\bibitem{} Burstein D., Heiles C., 1982, AJ 1165, 87

\bibitem{} Colless M., 1995, AJ 109, 1937

\bibitem{} Colless M., 1989, MNRAS 237, 799

\bibitem{} de Theije P.A.M., Katgert P., 1998, A\&A submitted (paper VI)

\bibitem{} Durret F., Forman W., Gerbal D., Jones C., Vikhlinin A., 1998, 
A\&A accepted

\bibitem{} Feldman H.A., Watkins R., 1994, ApJ 430, 17

\bibitem{} Heydon-Dumbleton N.H., Collins C.A., MacGillivray H.T., 1989,
MNRAS 238, 379

\bibitem{} Hudson M., Ebeling H., 1997, ApJ 479, 621

\bibitem{} Isobe T., Feigelsen E.D., Akritas M.G., Babu G.J., 1990, ApJ
364, 104

\bibitem{} Jones B.J.T., Mazure A., 1993, to app. in Measuring, Mapping
and Modelling the Universe ($http://www.tac.dk/bjones/$)

\bibitem{} Katgert P., Mazure A., den Hartog R., et al., 1998, A\&A
accepted (K98)

\bibitem{} Katgert P., Mazure A., Perea J., et al., 1996, A\&A 310, 8

\bibitem{} Lauer T.R., Postman M., 1994, ApJ 425, 418

\bibitem{} Lobo C., Biviano A., Durret F., et al., 1996, A\&A 317, 385

\bibitem{} Lubin P., Villela T., 1986, in Galaxy Distances and Deviations
from Universal Expansion, ed. B.F. Madore and R.B. Tully, Dordrecht (Reidel
p.169)

\bibitem{} Lumsden S.L., Collins M.A., Nichol R.C., Eke V.R., Guzzo L.,
1997, MNRAS 290, 119 (L97)

\bibitem{} Mazure A., Katgert P., den Hartog R., et al., 1996, A\&A 310, 31 

\bibitem{} Merrifield M.R., Kent S.M., 1989, AJ 98, 351

\bibitem{} Perlmutter S., Gabi S., Goldhaber G., et al., 1997, ApJ 483, 565

\bibitem{} Schaeffer R., Maurogordato S., Cappi A., Bernardeau F., 1993, MNRAS
263, L21

\bibitem{} Schechter P.L., 1976, ApJ 203, 297

\bibitem{} Trentham N., 1998, Astro-ph (9804013)
           
\bibitem{} Valotto C.A., Nicotra M.A., Muriel H., Lambas D.G., 1997, ApJ
479, 90 (V97)

\end{thebibliography}
\end{document}